\documentclass[useAMS,usenatbib,usegraphicx]{mn2e}

\title[Predicting the frequencies of diverse exo-planetary systems]
{Predicting the frequencies of diverse exo-planetary systems}
\author[J. S. Greaves, D. A. Fischer, M. C. Wyatt, C. A. Beichman \& G. 
Bryden]{J.S. Greaves$^{1}$\thanks{E-mail: jsg5@st-andrews.ac.uk 
(JSG)}, D.A. Fischer$^{2}$, M.C. Wyatt$^{3}$, C.A. Beichman$^{4,5}$  
and G. Bryden$^{4}$\\
$^{1}$Scottish Universities Physics Alliance, Physics and Astronomy, 
University of St. Andrews, North Haugh, St Andrews, Fife KY16, UK \\
$^{2}$Department of Astronomy, University of California at Berkeley, 601 
Campbell Hall, Berkeley, CA 94720, USA \\
$^{3}$Institute of Astronomy, Madingley Rd, Cambridge CB3 0HA, UK \\
$^{4}$Michelson Science Center, California Institute of Technology,  
M/S 100-22, Pasadena, CA 91125, USA \\
$^{5}$Jet Propulson Laboratory, 4800 Oak Grove Drive, Pasadena, CA 
91109, USA.
}

\begin{document}

\date{Accepted 2007. Received 2007; in original form 2006}

\pagerange{\pageref{firstpage}--\pageref{lastpage}} \pubyear{2006}

\maketitle

\label{firstpage}

\begin{abstract}
Extrasolar planetary systems range from hot Jupiters out to icy
comet belts more distant than Pluto. We explain this diversity in a
model where the mass of solids in the primordial circumstellar disk
dictates the outcome. The star retains measures of the initial
heavy-element (metal) abundance that can be used to map solid masses
onto outcomes, and the frequencies of all classes are correctly
predicted. The differing dependences on metallicity for forming
massive planets and low-mass cometary bodies are also explained. By
extrapolation, around two-thirds of stars have enough solids to form 
Earth-like planets, and a high rate is supported by the first 
detections of low-mass exo-planets. 
\end{abstract}

\begin{keywords}
circumstellar matter -- planetary systems: protoplanetary discs --
planetary systems: formation
\end{keywords}

\section{Introduction}

Extrasolar planetary systems have largely been identified by a
change in the line-of-sight velocity in spectra of the host star,
the `Doppler wobble' method \citep[e.g.]{cumming}. This technique
detects inner-system gas-giant planets out to a few astronomical
units.  Contrasting larger-scale systems are those with `debris'
disks, rings of dust particles produced in comet collisions, whose
presence indicates that parent bodies exist at least up to a few
kilometres in size \citep{wd}. Most images of debris disks show
central cavities similar to that cleared by Jupiter and Saturn in
our own Solar System \citep{liou}, and also sub-structure attributed
to dust and planetesimals piled up in positions in mean motion
resonance with a distant giant planet \citep{wyatt}. 
\citet{chas-both} have discovered a few systems with both debris
disks and inner giants, linking these divergent outcomes. 

These various planetary systems could reflect different initial
states, in particular the quantity of planet-forming materials in
the circumstellar disk around the young host star. In core-growth
models \citep[e.g.]{pollack,hubickyj}, a large supply of refractory 
elements (carbon, iron, etc.) should promote rapid growth of 
planetesimals that can then amalgamate into planetary cores. If gas 
still persists in the disk, the core can attract a thick atmosphere 
and form into a gas giant planet; the disk is also viscous so that 
the planet tends to migrate inwards \citep{nelson}. For sparse 
refractories, however, only small comets up to Neptune-like `ice 
giants' may have formed when the gas disperses; \citet{plstars} 
suggested that such systems will be characterised by planetesimal 
collisions and hence debris emission. 

Here we classify these outcomes from most to least successful, and
postulate that the {\it dominant} agent is the initial mass in
refractories. We aim to test whether one underlying property has a
highly significant effect on the outcome, and so intentionally ignore
other properties that could affect the planetary system formed around
an individual star. Such properties include the disk size, geometry,
composition and lifetime, as well as the stellar accretion rate,
emission spectrum and environment. Stochastic effects are also
neglected, such as inwards migration caused by inter-planet
encounters \citep{marzari} and debris brightening after collisions of
major planetesimals \citep{dominik}. Such factors are beyond the
scope of our simple model, but are important for detailed
understanding of the formation of particular kinds of planetary
system. 

\begin{figure}
\label{fig1}
\includegraphics[width=57mm,angle=270]{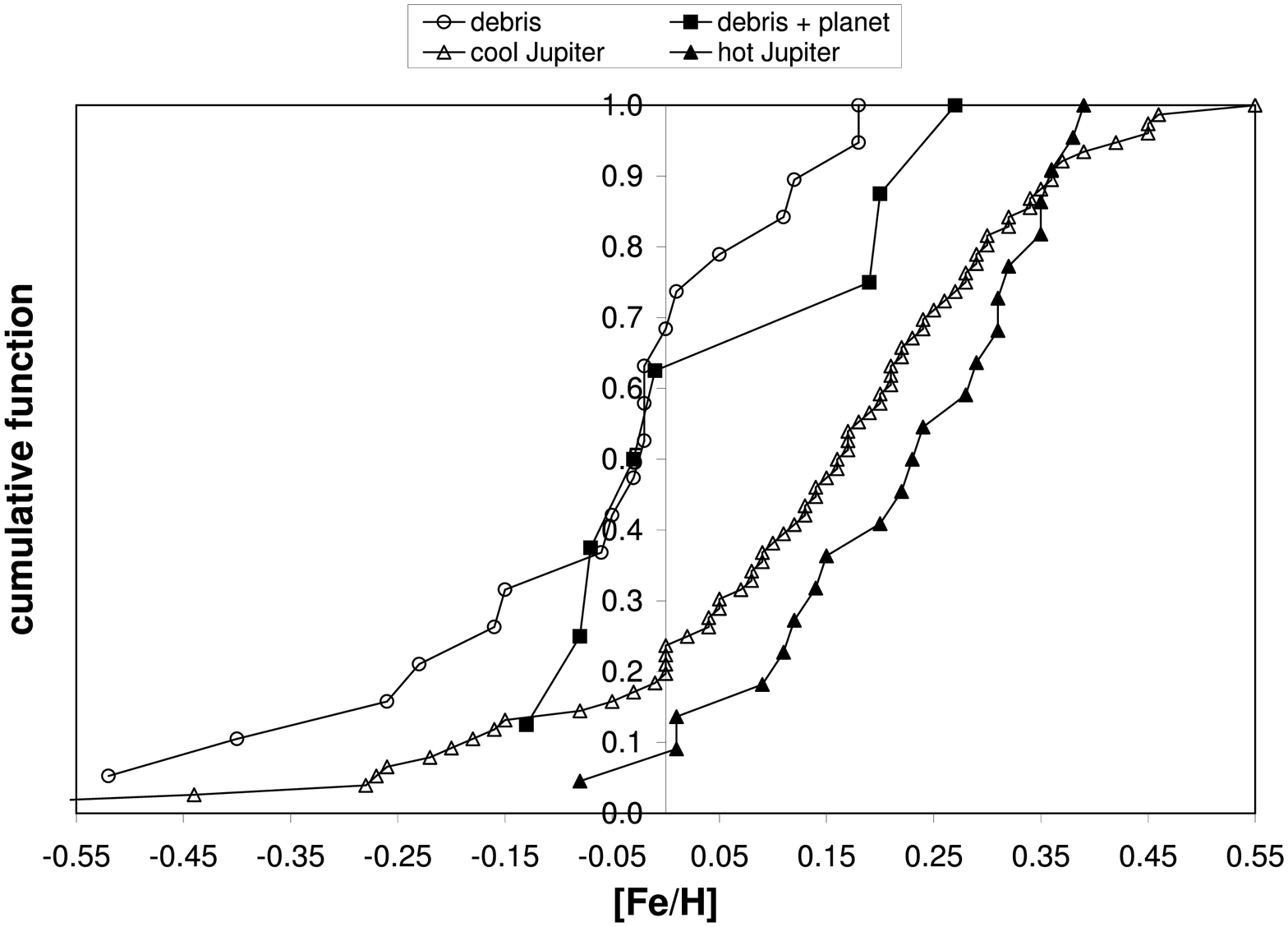}
\caption{Cumulative functions of [Fe/H] for the host stars of hot
Jupiters (at $\leq$ 0.1 AU), cool Jupiters ($>$ 0.1 AU),  
debris-and-planet systems, and debris disks only. One cool Doppler 
companion with [Fe/H] of --0.65 is not shown, as it is suspected to 
be above planetary mass \citep{fv05}.
}
\end{figure}

\section{Hypothesis}

The hypothesis explored here is that the mass of solid elements in a
primordial circumstellar disk can be quantified, and linked to an
outcome such as a detectable debris disk or radial velocity planet.
The only piece of relevant `relic' information for a particular star
is the metallicity, quantified by [Fe/H], the logarithmic abundance of
iron with respect to hydrogen and normalised to the Solar value, and
this quantity is taken here to track refractories in general. In
combination with a distribution of total (gas-plus-dust) masses of
primordial disks, masses of solids can then be inferred
statistically. Our basic hypothesis is that the metallicity is a relic
quantity originally common to the young star and its disk, and that
higher values of the trace refractory iron should correlate with more
effective planet growth. 

It is well-known that gas giant detection rates rise at higher [Fe/H]
\citep[e.g]{fv05}, while \citet{gfw} have noted that debris detection
is essentially independent of metallicity. \citet{robinson} have shown
that the growth of gas giant planets can be reproduced in simulations
taking disk mass and metal content into account. Here we extend such
ideas to include systems with Doppler planets, with debris disks, and
with both phenomena. 

Metal-based trends are now identified (Figure 1) for four
outcomes\footnote{The separation of hot and cool Jupiters in [Fe/H] has
been previously noted as having modest significance
\citep{santos,sozzetti}; a K-S test here gives P=0.35 for the null
hypothesis using the Fig.~1 data (errors in [Fe/H] of typically 0.025
dex). The overlap of the two distributions is in fact an integral part
of our model, where {\it two} factors contribute to the outcome.}
ranked from most to least successful: a) hot Jupiters, b) more distant
Doppler planets, with semi-major axis beyond 0.1~AU, c) systems with
both giant planets and debris, and d) stars with debris only. Here we
base `success' (implicitly, via core-accretion models) on a large
supply of refractories and so fast evolutionary timescales, with hot
Jupiters rapidly building a core, adding an atmosphere and migrating
substantially towards the star. In systems with progessively less
success: planets form more slowly so they migrate less over the
remaining disk lifetime (cool Jupiters); only some of the material is
formed into planets (planet-and-debris systems); and mainly
planetesimals are formed, perhaps with planets up to ice-giant size
(debris).  The 0.1~AU divide of hot/cool Jupiters is not
necessarily physical, and is simply intended to give reasonable source
counts. Notably, in systems with debris plus Jupiters, the planets
orbit at $>$ 0.5 AU, consistent with even lower success in migration. 

The plotted ranges of [Fe/H] shift globally to higher values with
more success, as expected -- however, the ranges also overlap
indicating that the fraction of metallic elements is not the only
relevant quantity. The hypothesis to be tested here is that the
underlying dominant factor is the {\it total mass} of metals in the
primordial disk. This mass is the product of the total mass M of
the disk (largely in H$_2$ gas) and the mass-fraction Z of
refractory elements. The range of Z for each outcome is broad
because disks of different M were initially present; thus, a
massive low-metal disk and a low-mass metal-rich disk could both
lead to the same outcome. 

\section{Model}

The distribution of the product M$_S \propto$ M Z was constructed from a 
mass distribution measured for primordial disks \citep{andrews} and 
our metal abundances measured in nearby Sun-like stars \citep{vf05}. 
With the canonical gas-to-dust mass ratio of 100 for present-day 
disks, the mass of solid material is then 
\begin{equation}
M_S = 0.01 M \times 10^{{\rm [Fe/H]}},
\end{equation}
with the modifying Z-factor assumed to be traced by iron. 
\citet{padgett} and \citet{james} find that the metallicities of 
present-day young star clusters are close to Solar (perhaps lower by 
$\sim$~0.1 dex), so the reference-point values of 100 for 
the gas-to-dust ratio and 0 for [Fe/H] are self-consistent to good 
accuracy. 

Our model has the following assumptions. 
\begin{itemize}
\item{There are contiguous bands of the M$_S$ distribution, 
corresponding to different planetary system outcomes. Any one disk has 
an outcome solely predicted by its value of M$_S$. The ranges of M 
and Z can overlap, as only the M$_S$ bands are required to be 
contiguous.}
\item{A particular outcome will arise from a set of M Z products, with 
the observable quantity being the metallicity ranging from Z(low) to
Z(high). This Z-range sets the M$_S$ bounds, as described below. } 
\item{For the stellar ensemble, all possible products are assumed
to occur, and to produce some outcome that is actually observed. 
There are no unknown outcomes (novel kinds of planetary system not 
yet observed).} 
\item{M and Z are taken to be independent, as the disk mass is
presumably a dynamical property of the young star-disk system and the
mass in solids is always a small component.}
\end{itemize}

\begin{table*}
 \caption{Results of the planetary system frequency calculations. The
ranges of solid mass M$_S$(low) to M$_S$(high) were determined based 
on the input data of the observed [Fe/H] range for each outcome. 
The total disk masses given are consistent with these two boundary 
conditions. The ranges of observed frequency include different surveys 
and/or statistical errors as discussed in the text. The predicted total 
is less than 100~\% because of a few `missing outcomes' (see text). 
The null set represents stars searched for both planets and debris with 
no detections. 
}
 \label{tab:list}
 \begin{tabular}{@{}cccccc}
  \hline
outcome		 & [Fe/H] range	  & solid mass			& total disk mass & predicted 	& observed      \\
		 &		  & (M$_{Jup}$ [M$_{\oplus}$])	& (M$_{Jup}$)	  & frequency	& frequency     
\\
  \hline
hot Jupiter	 & --0.08 to +0.39 & 1.7--5 [500--1600]		& 70--200  & 1 \%	& 2 $\pm$ 1 \%  \\
cool Jupiter	 & --0.44 to +0.40 & 0.24--1.7 [75--500]	& 10--200  & 8 \%	& 9 (5--11) \%  \\
planet \& debris & --0.13 to +0.27 & 0.10--0.24 [30--75]	& 5--30	   & 5 \%	& 3 $\pm$ 1 \%  \\
debris		 & --0.52 to +0.18 & 0.02--0.10	[5--30]		& 1--30	   & 16 \%	& 15 $\pm$ 3 \% \\
null		 & --0.44 to +0.34 & $<$ 0.02 [$< 5$]		& $<$ 15   & 62 \%	& $\sim 75$ \%  \\
  \hline
 \end{tabular}
\end{table*}

The M$_S$ bounds were determined using the minimum number of free
parameters. For a particular outcome, the observed Z-range is
assumed to trace the bounds of M$_S$, i.e. 
\begin{equation}
M_S(high) / M_S(low) = f \times Z(high) / Z(low)
\end{equation}
where $f$ is a constant, and this relation sets solid-mass bounds for
each outcome. Results presented here are mainly for $f = 1$, but more
complex relationships could exist, and other simple assumptions without
further variables could also have been made -- alternative forms
of Eq.~2 are explored in section 4.1
 
Absolute values for M$_S$ were derived iteratively, working downwards
from the most successful outcome. The upper bound for the
hot Jupiter band was derived from the highest disk mass observed and
highest metallicity seen for this outcome (under the assumption
that all M Z products are observed). Other outcomes were then
derived in turn assuming $f = 1$, and working down in order of success,
with each lower bound M$_S$(low) setting the upper bound M$_S$(high)
for the next most successful outcome. Both M and Z have log-normal
distributions, and 10$^6$ outcomes were calculated, based on 1000
equally-likely values of log(M) combined with 1000 equally-probable
values of log(Z), i.e. [Fe/H]. 

\subsection{Input data}

The M-distribution is taken from a deep millimetre-wavelength survey
for dust in disks in the Taurus star-forming region by
\citet{andrews}. \citet{nuernberger} have found similar results from
(less deep) surveys of other regions, so the Taurus results are taken
here to be generic, under the simplification that local environment
is neglected. The total disk masses M were found from the dust masses
multiplied by a standard gas-to-dust mass ratio of 100. The mean of
the log-normal total disk masses is 1 Jupiter mass (i.e. log M = 0
in M$_{Jup}$ units) with a standard deviation of 1.15 dex, and
detections were actually made down to $-0.6 \sigma$.  The
Z-distribution is from the volume-limited sample from \citet{fv05} of
main sequence Sun-like stars (F, G, K dwarfs) within 18 pc.  These
authors also list 850 similar stars out to larger distances that are
being actively searched for Doppler planets. In the 18 pc sample, the
mean in logarithmic [Fe/H] is --0.06 and the standard deviation is
0.25 dex. 

Both M and Z distributions have an upper cutoff at approximately the
$+2 \sigma$ bound: for M this is because disks are less massive than
$\sim 20$~\% of the star's mass (presumably for dynamical stability),
and for Z because the Galaxy has a metal threshold determined by
nucleosynthetic enrichment by previous generations of stars. The
upper cutoff for M is 200 M$_{Jup}$ for `classical T Tauri' stars,
and for Z the adopted cutoff in [Fe/H] is +0.40 from the 18 pc sample
(with a few planet-hosts of [Fe/H] up to 0.56 found in larger
volumes). 

\subsection{Observed frequencies}

The Doppler-detection frequencies quoted in Table 1 are mostly from
the set of 850 uniformly-searched stars with [Fe/H] values. The
statistics for hot Jupiters range from 16/1330 = 1.2~\% in a
single-team search \citep{marcy} up to 22/850 = 2.6~\% in the uniform
dataset \citep{fv05}.  For cool Jupiters (outside 0.1 AU semi-major
axis), the counts are 76/850 = 8.9~\% \citep{fv05}, with \citet{marcy}
finding a range from 72/1330 = 5.4~\% within 5 AU up to an
extrapolation of 11~\% within 20 AU. The upper limit is based on
long-term trends in the radial velocity data, and these as-yet
unconfirmed planetary systems could have [Fe/H] bounds beyond those 
quoted here. 

The debris counts are based on our surveys with Spitzer. The
debris-only statistics \citep{chas-06,bryden} comprise 25 Spitzer
detections out of 169 target stars in unbiased surveys, i.e. 15 $\pm$
3~\% with Poissonian errors\footnote{Figure~1 also includes a few
prior-candidate systems
(www.roe.ac.uk/ukatc/research/topics/dust/identification.html)
confirmed by Spitzer programs.}. For planet-plus-debris systems, the
3~\% frequency is estimated from 6/25 detections (24 $\pm$ 10~\%)
among stars known to have Doppler planets \citep{chas-both},
multiplied by the 12~\% total extrapolated planet frequency
\citep{marcy}. In Table 1, the planet-only rates should strictly sum
to only up to 9~\% if this estimate of 3~\% of planet-and-debris
systems is subtracted.

The null set quoted has an [Fe/H] range derived from our Spitzer
targets where no debris or planet has been detected. The M$_S$(low)
bound for the null set has been set to zero rather than the formal
limit derived from Z(low)/Z(high), as lower values of M$_S$(low)
presumably also give no presently observable outcome. 

\section{Results}

The predicted frequencies (Table 1) agree closely with the
observed rates in {\it all} outcome categories. This is remarkable
when very different planetary systems are observed by
independent methods, and ranging in scale from under a tenth to
tens of AU. The good agreement suggest that solid mass may indeed be 
the dominant predictor of outcome. 

The model is also rather robust. In particular, because the
M-distribution is an independent datum, obtaining a good match of
predicted and observed frequencies is not inevitable. For example,
artificially halving the standard deviation of the M distribution
would yield far too many stars with detectable systems (75~\%), 
and many more cool Jupiters than hot Jupiters (around 20:1 instead 
of 5:1). The model is also reasonably independent of outlier
data points. For example, adding in the low-metallicity system
neglected in Figure~1 would raise the prediction for cool Jupiters
from 8~\% to 12~\%, or extending the upper Z-cutoff
to the [Fe/H] of the distant Doppler systems would raise this
prediction to 11~\%. However, the resulting effects on the
less-successful system probabilities are less than 1~\%.  This 
suggests that although small number statistics may affect
the predictions, the results would not greatly differ if large
populations were available, that could be described by statistical
bounds rather than minimum- to maximum-[Fe/H]. 

The model also accounts for nearly all outcomes, as required if each
M$_S$ is to result in only one planetary system architecture. For a
few M$_S$ products, one of the outcomes would be expected except that
the Z value involved lies outside the observed ranges. These
anomalous systems sum to 9~\% (hence the Table 1 predictions add 
to $< 100$~\%). These `missing' systems are predominantly debris and
debris-plus-planet outcomes. The latter class has the smallest number
of detections (Figure~1), and more examples are likely to be
discovered with publication of more distant Doppler planets. 

A further check is that the disk masses are realistic, in
terms of producing the observed bodies. Doppler systems are predicted
here to form from 5-200 Jupiter masses of gas in the disk, enough to
make gas giant planets. Also, the M$_S$ values of 30-1600
Earth-masses could readily supply the cores of Jupiter and Saturn
(quoted by \citet{saumon} as $\approx 0-10$ and $\approx 10-20$ 
M$_{\oplus}$ respectively) or the more extreme $\sim 70$ M$_{\oplus}$
core deduced for the transiting hot Jupiter around HD~149026
\citep{sato}. Similarly, populations of colliding bodies generating
debris have been estimated at $\sim$~1--30 Earth-masses
\citep{wd,tauceti}, a quantity that could reasonably be produced from
5--75 Earth-masses in primordial solids (Table 1). 

To make Jupiter analogues within realistic timescales, core growth
models \citep[e.g]{hubickyj} need a few times the Minimum Mass
Solar Nebula, which comprised approximately 20 M$_{Jup}$
\citep{davis}. The Solar System itself would thus have contained a
few times 0.2 M$_{Jup}$ in solids, of which 0.15 M$_{Jup}$ (50
M$_{\oplus}$) has been incorporated in planetary cores. These
primordial disk masses would place the Solar System in the cool
Jupiter category, and this is in fact how it would appear
externally (the dust belt being more tenuous than in detected
exo-systems). 

\subsection{Alternative models}

Equation (2) was further investigated with $f \neq 1$. Reasonable
agreement with the observations was obtained only for $f$ in a narrow
range, $\approx 0.8-1.1$. For higher $f$, systems with debris are
over-produced, while for lower $f$ there are too few Doppler
detections. This suggests that the hypothesis that the Z-range
directly traces the M$_S$ range for the outcome to occur is close to
correct, although not well understood. Qualitatively, there is a
locus of points in M, Z parameter space that lie inside the
appropriate M$_S$ bounds for an outcome, and at some mid-range value
of M it is likely that all the Z(low) to Z(high) values are
appropriate, and so this traces the product M$_S$(high) to M$_S$(low)
(provided no more extreme values of Z are suitable at the extrema of
M). Z seems to have the most constrictive effect on outcome because
the distribution is much narrower than that of M, and thus if M
changes by a large amount, there is no corresponding value of Z than
can preserve a similar $M_S$. Simulations of planetesimal growth as a
function of mass of solids in the disk are needed to further explore
why the ranges of Z and $M_S$ match so closely. 

One even simpler model was tested, with a constant range of
$M_S(high) / M_S(low)$ for each observable outcome. Assuming that
$\sim 1$~M$_{\oplus}$ of solids is needed for the minimum detectable
system (planetesimals creating debris), and that the most massive
disks contain 1600~M$_{\oplus}$ of solids (from the maximum M and Z),
then this mass range can be divided into equal parts for the four
observable outcomes, with a range of log-M~=~0.8 for each. This
simple model fails, in particular greatly over-producing
debris-and-planet systems at 12~\%. Thus, it seems that the Z-range
does in fact contain information on the outcomes.

\subsection{Distributions within outcomes}

The dependences on metallicity of planet and debris detections
arise naturally in the model.  Doppler detections are strongly
affected by metallicity: \citet{fv05} find that the
probability is proportional to the square of the number of iron
atoms. Our model predicts P(hot Jupiter) $\propto$ N(Fe)$^{2.7}$
and P(cool Jupiter) $\propto$ N(Fe)$^{0.9}$, above 80~\%- and
30~\%-Solar iron content respectively. The observed exponent of 2
for all planets is intermediate between the two model values, and
as predicted by \citet{robinson}, a steeper trend for short-period
planets is perhaps seen, at least at high metallicities (Figure~1).
These dependences arise because large solid masses are needed for
fast gas giant formation with time for subsequent migration, and so
when [Fe/H] is low a large total disk mass is required, with
rapidly decreasing probability in the upper tail of the
M-distribution. 

In contrast, the model predicts a weak relation of P(debris)
$\propto$ N(Fe)$^{0.2}$ (for 30-160~\%-Solar iron), agreeing
with the lack of correlation seen by \citet{bryden}. As fewer solids
are needed to make comets, a wide variety of parent disks
will contain enough material, so little metal-dependence is expected. 
For example (Table~1), M$_S$(low) can result from a 6 M$_{Jup}$ disk
with [Fe/H] of --0.5 or a 1.2 M$_{Jup}$ disk with [Fe/H] of +0.2.  As
these masses are both central in the broad M-distribution, they have
similar probability, and so the two [Fe/H] values occur with similar
likelihood. 

\section{Discussion}

The excellent reproduction of observed frequencies and dependencies on
metallicity both suggest that the method is robust. Hence, rather
surprisingly, the original search for a single parameter that has a
dominant effect on outcome has succeeded. Under the assumption that the
metallicity ranges track the different outcomes, the primordial solid
mass of the circumstellar disk is identified as this parameter. 

One implication is that for an ensemble of stars of known metallicity,
the proportions of different kinds of planetary system may be
predicted, without the need for detailed models of individual disks.
As main-sequence stars retain no relic information on their primordial
disk masses, this is a very useful result, leading to estimates of the
frequency and variety of planetary systems, for example among nearby
Sun-analogues of interest to planet-detection missions. 

\begin{figure}
\label{fig2}
\includegraphics[width=38mm,angle=270]{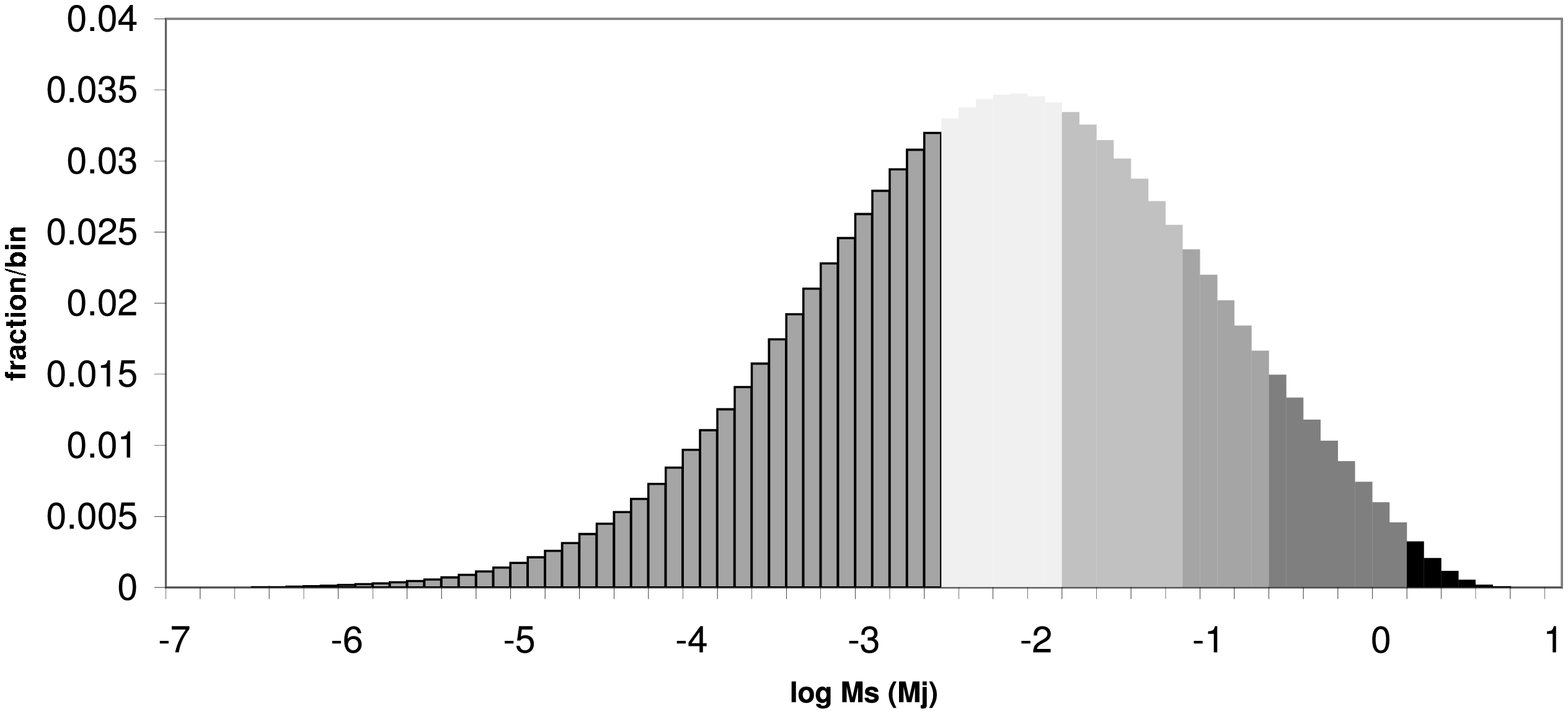}
\caption{Distribution of solid masses. The shaded areas represent 
(from right to left) systems producing hot Jupiters, cool Jupiters, 
gas giants plus debris disks, debris only, and disks with 1 Earth-mass 
or more of solids. Disks further to the left have no presently 
predicted observable outcome.
}
\end{figure}

The model may also be used to examine other planetary
system regimes that have just opened up to experiment. For example,
transiting hot Jupiters have not been detected in globular clusters,
although the stellar density allows searches of many stars. In
47~Tuc, no transits were found amongst $\sim 20,000$ stars, although
7 detections would be expected based on the typical hot Jupiter
occurrence rate \citep{weldrake}. Assuming these old stars formed
with disks following the standard M-distribution, but with
metallicities of only one-fifth Solar in a narrow range with $\sigma
\approx$0.05 dex \citep{caretta}, then the model predicts that less
than 1 in 10$^6$ disks can form a hot Jupiter -- the solid masses are
too small for fast planet growth and subsequent migration. 

Finally, an upper limit can be estimated for the number of young disks
that could form an Earth-mass planet. The maximum fraction (Figure~2)
is set by disks of M$_S \geq 1$~M$_{\oplus}$, summing to two-thirds of
stars. Thus one in three stars would not be expected to host any
Earth-analogue, and irrespective of metallicity since the occurrence of
this minimum mass is rather flat with P $\propto$ N(Fe)$^{0.25}$.
However, if the upper two-thirds of disks may be able to form
terrestrial planets, this would agree with the large numbers predicted
by the simulations of \citet{idalin}, and also with the first planetary
detections by the microlensing method. Two bodies of only around 5 and
13 Earth-masses have been detected around low-mass stars, and
\citet{gould} estimate a frequency of 0.37 (--0.21, +0.30) in this
regime of icy planets orbiting at $\sim 3$~AU. Our model finds that
40~\% of stars have disks with M$_S$ greater than 5~M$_{\oplus}$, so
the minimum materials to form such planets are present at about the
observed frequency. While very preliminary, this result supports the
prediction that many stars could have low-mass planets. 

\section{Conclusions}

We tested the hypothesis that a single underlying parameter could have
the dominant effect on the outcome of planetary system formation from
primordial circumstellar disks. An empirical model with the mass of
solids as this parameter produces a very good match to the observed
frequencies and to the dependence on host-star metallicity, when the
metallicity range within each outcome is used to estimate the
solid-mass range. This model may be very useful for making statistical
predictions of planetary system architectures for ensembles of stars of
known metallicity, such as nearby Solar analogues of interest to
exo-Earth detection missions.

\section*{Acknowledgments}
JSG thanks PPARC and SUPA for support of this work. We thank the 
referee, Philippe Thebault, for comments that greatly helped the 
paper.

\bsp

\label{lastpage}

\end{document}